\documentclass[pdflatex,sn-mathphys-num]{sn-jnl}

\newcommand{\aar}{Astron. Astrophys. Rev.} 
\newcommand{\aj}{Astron. J.}   
\newcommand{\apj}{Astrophys. J.}   
\newcommand{\apjl}{Astrophys. J. Lett.}   
\newcommand{\apjs}{Astrophys. J. Suppl. Ser.}   
\newcommand{\aap}{Astron. Astrophys.}   
\newcommand{\mnras}{Mon. Not. R. Astron. Soc.}   
\newcommand{\nat}{Nature} 
\newcommand{\nastro}{Nat. Astron.} 
\newcommand{\pasp}{Publ. Astron. Soc. Pac.}   
\newcommand{\uni}{Universe} 
\usepackage{graphicx}%
\usepackage{placeins}
\usepackage{multirow}%
\usepackage{amsmath,amssymb,amsfonts}%
\usepackage{amsthm}%
\usepackage{mathrsfs}%
\usepackage[title]{appendix}%
\usepackage{xcolor}%
\usepackage{textcomp}%
\usepackage{manyfoot}%
\usepackage{booktabs}%
\usepackage{algorithm}%
\usepackage{algorithmicx}%
\usepackage{algpseudocode}%
\usepackage{listings}%
\usepackage{newfloat}
\usepackage{etoolbox}
\renewcommand{\tablename}{Extended Data Table}
\DeclareFloatingEnvironment[fileext=lof,placement=ht,
    name=Extended Data Figure]{extfigure}
\theoremstyle{thmstyleone}%
\theoremstyle{thmstyletwo}%

\theoremstyle{thmstylethree}%

\newcommand{\orcidlink}[1]{%
\href{https://orcid.org/#1}{\textsuperscript{\textcolor{green!50!black}{\small\textbullet}}}%
}
\begin{document}

\title[Article Title]{Ram-pressure stripping caught in action in a young cluster at $z=2.51$}

\author[1,2]{\fnm{Ke} \sur{Xu} \orcidlink{0000-0002-8046-984X}}

\author*[1,2]{\fnm{Tao} \sur{Wang}\orcidlink{0000-0002-2504-2421}}\email{taowang@nju.edu.cn}

\author[3]{\fnm{Emanuele} \sur{Daddi}\orcidlink{0000-0002-3331-9590}}

\author[3]{\fnm{David} \sur{Elbaz}\orcidlink{0000-0002-7631-647X}}

\author[1,2]{\fnm{Hanwen} \sur{Sun}\orcidlink{0009-0007-0241-0213}}

\author[1,2]{\fnm{Longyue} \sur{Chen}\orcidlink{0009-0001-7925-4876}}

\author[1,2]{\fnm{Qiaoyang} \sur{Hao}\orcidlink{0009-0000-2222-3713}}

\author[4]{\fnm{Raphael} \sur{Gobat}\orcidlink{0000-0003-0121-6113}}

\author[5]{\fnm{Anita} \sur{Zanella}\orcidlink{0000-0001-8600-7008}}

\author[6,7]{\fnm{Daizhong} \sur{Liu}\orcidlink{0000-0001-9773-7479}}

\author[8]{\fnm{Mengyuan} \sur{Xiao}\orcidlink{0000-0003-1207-5344}}

\author[9]{\fnm{Renyue} \sur{Cen}\orcidlink{0000-0001-8531-9536}}

\author[10]{\fnm{Tadayuki} \sur{Kodama}\orcidlink{0000-0002-2993-1576}}

\author[11,12]{\fnm{Kotaro} \sur{Kohno}\orcidlink{0000-0002-4052-2394}}

\author[1,2]{\fnm{Tiancheng} \sur{Yang}\orcidlink{0009-0008-4971-035X}}

\author[1,2,13]{\fnm{Can} \sur{Xu}\orcidlink{0000-0002-8437-6659}}

\author[1,2]{\fnm{Zhi-Yu} \sur{Zhang}\orcidlink{0000-0002-7299-2876}}

\author[1,2]{\fnm{Luwenjia} \sur{Zhou}\orcidlink{0000-0003-1687-9665}}

\author[14,15]{\fnm{Francesco} \sur{Valentino}\orcidlink{0000-0001-6477-4011}}

\affil[1]{School of Astronomy and Space Science, Nanjing University, Nanjing, China}
\affil[2]{Key Laboratory of Modern Astronomy and Astrophysics, Nanjing University, Nanjing, China}
\affil[3]{Université Paris-Saclay, Université Paris Cité, CEA, CNRS, AIM, Gif-sur-Yvette, France}
\affil[4]{Instituto de Física, Pontificia Universidad Católica de Valparaíso, Valparaíso, Chile}
\affil[5]{Istituto Nazionale di Astrofisica (INAF), Padua, Italy}
\affil[6]{Purple Mountain Observatory, Chinese Academy of Sciences, Nanjing, China}
\affil[7]{State Key Laboratory of Radio Astronomy and Technology, Purple Mountain Observatory, Chinese Academy of Sciences, Nanjing, China}
\affil[8]{Department of Astronomy, University of Geneva, Versoix, Switzerland}
\affil[9]{Centre for Cosmology and Computational Astrophysics, Institute for Advanced Study in Physics and Institute of Astronomy, School of Physics, Zhejiang University, Hangzhou, China}
\affil[10]{Astronomical Institute, Tohoku University, Sendai, Japan}
\affil[11]{Institute of Astronomy, School of Science, The University of Tokyo, Tokyo, Japan}
\affil[12]{Research Centre for the Early Universe, School of Science, The University of Tokyo, Tokyo, Japan}
\affil[13]{Kavli Institute for the Physics and Mathematics of the Universe (WPI), The University of Tokyo, Chiba, Japan}
\affil[14]{Cosmic Dawn Center (DAWN), Copenhagen, Denmark}
\affil[15]{DTU Space, Technical University of Denmark, Kongens Lyngby, Denmark}



\abstract{Galaxy clusters in the local Universe are dominated by massive quiescent galaxies with old ages, formed at high redshifts. Whether
their quenching is driven by internal processes or environmental effects is a matter of debate that has been challenging to resolve
due to the lack of observations during their peak formation epoch. Here we report clear evidence from the Atacama Large Millimeter/submillimeter Array of extended and elongated gas tails in five galaxies in a forming cluster at $z = 2.51$. The single-tailed gas distributions, which extend notably beyond the stellar emission probed by JWST in galaxies that are relatively isolated and lack signatures of mergers or interactions (features that are very uncommon in the field), provide evidence of ram-pressure stripping. These very distant confirmed cases of ram-pressure stripping highlight the critical role of environmental effects in gas removal at high redshifts, an often-overlooked quenching pathway.}

\maketitle
\section*{Main}\label{sec1}

Direct evidence of the impact of environmental effects on cluster galaxies during their major formation epoch, $z \approx 2$--4, has so far been lacking. Most structures discovered at these epochs are either too young and extended with their halo masses far below typical clusters, or already relatively mature with a dominant population of quiescent galaxies. 
A novel type of structures has recently been discovered at $z\gtrsim2$~
\cite{Miller:2018,Capak:2011,Koyama:2013,Miley+2006,Dannerbauer+2014,Cucciati+2018,Oteo+2018,Shimakawa+2018,Daddi+2021}, that exhibits high overdensities of galaxies (and high halo mass) in the core but is dominated by star-forming galaxies. 
Among these structures, CLJ1001~\cite{Wang+16,Champagne:2021} at $z=2.51$ with a total mass close to a mature cluster ($M \sim 0.5-1 \times 10^{14}~ M_{\odot}$)~\cite{Wang+16}, could serve as a crucial bridge between more extended protoclusters and fully formed galaxy clusters. 
The abundance of massive star-forming galaxies (SFGs) and X-ray detections indicating the forming hot halo~\cite{Wang+16} make CLJ1001 an ideal laboratory to investigate whether and how the dense environment affects massive galaxy formation.

CLJ1001 harbours 14 SFGs with $\log(M_\star/{M}_\odot)>10.5$ within the central 340~kpc, most of which are expected to transition into quiescent galaxies soon, as indicated by their low gas fraction\cite{Wang+18}, compact sizes\cite{Xu+2023}, and a top-heavy galaxy stellar mass function\cite{Sun:2024}. 
To probe the role of dense environments in the evolution of these massive cluster SFGs, extensive multiwavelength observations have been conducted (Methods). 
In particular, the CO(3-2) observations from  the Atacama Large Millimeter/submillimeter Array (ALMA), combined with high-resolution imaging from the James Webb Space Telescope (JWST)/Near-Infrared Camera (NIRCam) from the COSMOS-Web Survey (a JWST Cycle 1 Treasury programme), enable studies of the spatial distributions of both molecular gas and stars, which are among the most sensitive probes of various environmental effects. 

\section*{Results}\label{sec2}

\begin{figure} 
\centering
\includegraphics[width=0.95\textwidth]{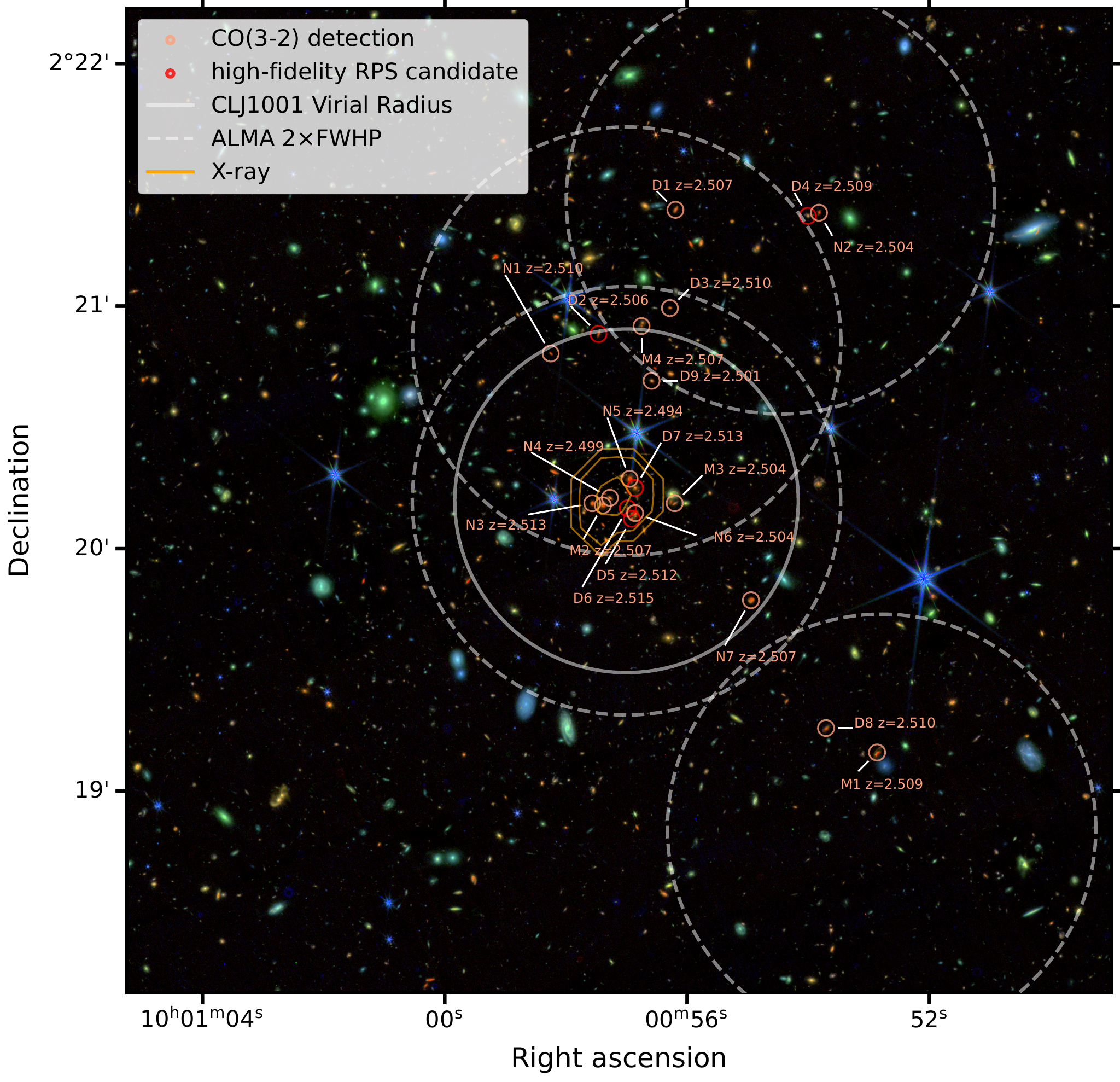}
\caption{\textbf{The sky image of CLJ1001.} CO(3-2) detections are shown in orange circles marked with the associated IDs and redshifts in orange text on the JWST/NIRCam RGB image (R: F444W, G: F277W and B: F150W). 
Red circles show the high-fidelity RPS sample.
The solid white circle shows the virial radius of CLJ1001 and the dashed white circles show the coverage of four ALMA pointings at 2$\times$ full-width at half-power (FWHP). The orange contours plot the X-ray detection from ref.~\cite{Wang+16}, increasing as [0.1,~0.5,~0.9]$\times$ the peak value.}\label{fig1}
\end{figure}

The CO(3-2) observations cover  member galaxies in CLJ1001 both in the centre and in the outskirts, up to twice the virial radius (throughout this work, we approximate the virial radius as the radius within which the mean density is 200 times the critical density of
the Universe, denoted $R_{\rm 200c}$), with a resolution of $1.09^{\prime\prime} \times 0.98^{\prime\prime}$ (and $1.02^{\prime\prime}\times0.93^{\prime\prime}$ for the lower single pointing; corresponding to the physical scale of $\sim$ 8~kpc at $z = 2.51$). In total, 20 member galaxies are clearly detected in CO(3-2) observations (information summarized in Extended Data Table 1). The spatial distributions of these CO(3-2)-detected members within CLJ1001 are shown in Fig.~\ref{fig1}. 

The reduced JWST/NIRCam imaging~\cite{Sun:2024} represents one of the first rest-frame near-infrared observations of cluster galaxies at $z > 2$ and covers most areas of CLJ1001, with four exposures in four filters (F115W, F150W, F277W and F444W), reaching 0.14$''$ spatial resolution in F444W and 5$\rm\sigma$ point-source depths of stellar mass of $\sim$ 10$^{8.4}$~$M_\odot$ at the redshift of CLJ1001 for a stellar population formed in a burst at $z\approx$ 20.
\subsection*{Galaxies exhibiting disturbed gas morphologies}
\begin{figure}
\centering
\includegraphics[width=0.95\textwidth]{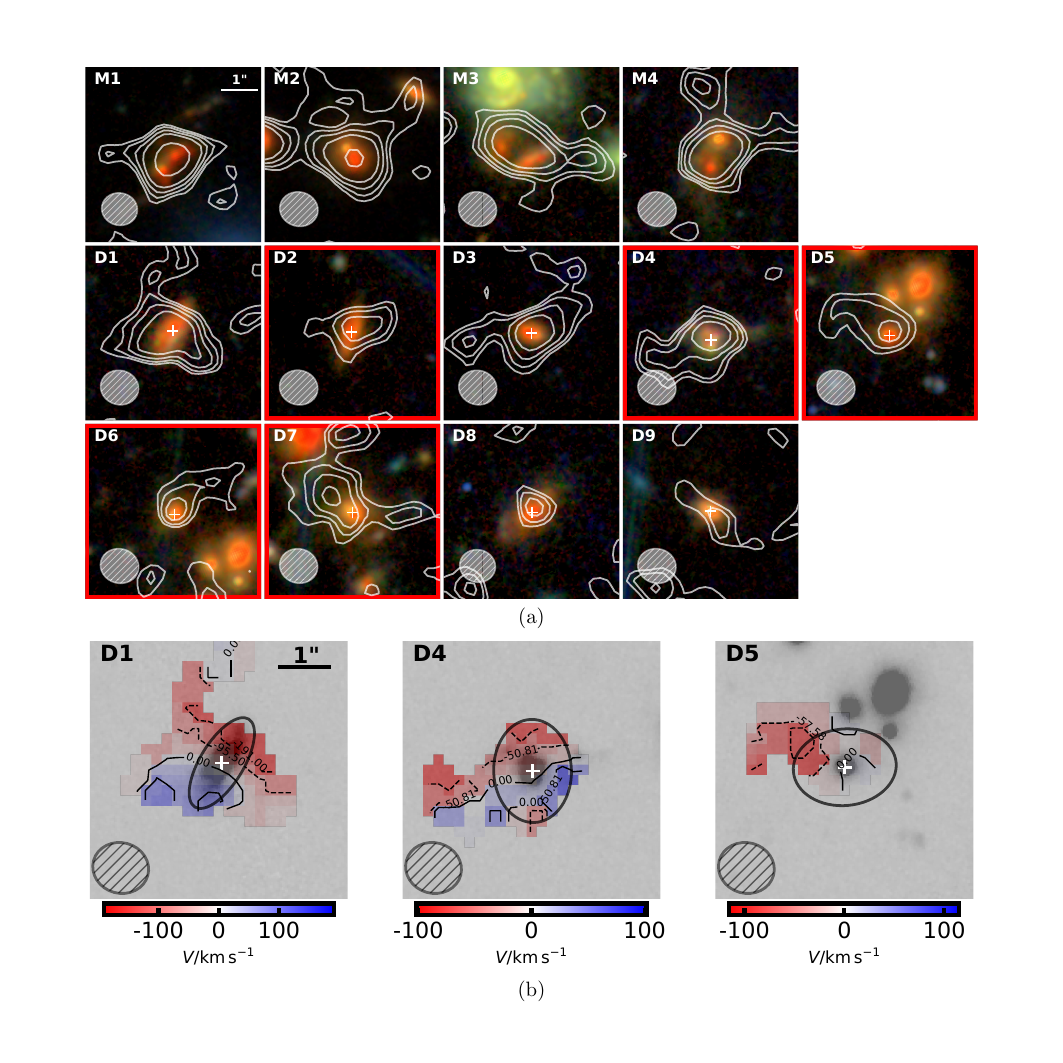}
\caption{{\bf CO(3-2) moment-0 maps on the false colour JWST/NIRCam images for the mergers and disturbed galaxies and some example moment-1 maps on the F444W image.}
(a), CO(3-2) distributions for the galaxies indicated (white contours; the levels are $\sqrt{2}$, 2, 2$\sqrt{2}$, 4, 8, 16 times the root mean square (r.m.s.)). The background images are false-colour NIRCam images made with F444W (R), F277W (G) and F150W (B) images. The red boxes highlight the high-fidelity gas-stripping cases. All panels are $5''\times5''$ in size. The scale bar in the M1 map applies to all maps in (a). (b), Moment-1 maps for D1, D4 and D5. The contours and values denote the velocity gradients and velocities ($V$). The white plus symbols mark the centres of the stellar components. The black ellipses at the centres show the galaxy morphology from \textsc{PetroFit}. The ellipses in the lower left corners of each map denote the synthesized beam of CO(3-2) observations. All panels are $5''\times5''$ in size. The scale bar in the left map applies to all maps in (b).}\label{fig2}
\end{figure}

We compare the spatial distribution of molecular gas and stellar components for these member galaxies via their CO(3-2) intensity maps (moment-0) and JWST/NIRCam images.
As shown in Fig.~\ref{fig2}, some of these galaxies show a more extended CO(3-2) distribution than their stellar components (in Extended Data Figure 1 we also show the CO(3-2) distribution with a high-resolution data cube). While some of them are clearly mergers (4/20, M1 to M4) with peaks in gas emission between two galactic nuclei, the majority (9/20, disturbed galaxies: D1 to D9) are relatively isolated galaxies with asymmetric CO(3-2) structures. They generally feature gas elongation beyond the stellar component (D1 to D6) or a gas disk offset from the galaxy centre (D7 to D9).
The median stellar mass of the disturbed galaxies is smaller than that of the normal galaxies and mergers ($\log(M_\star/{M_\odot})=10.55$ versus $11.02$; shown in Extended Data Figure 2), implying a less gravitationally bound gaseous component in disturbed galaxies.
Among these disturbed galaxies, D2, D4, D5, D6 and D7 exhibit extended gas tails in one direction. This is consistent with typical ram-pressure stripping (RPS) cases, and we classify them as the high-fidelity sample. D1, D3, D8 and D9 exhibit more complex or shrunken gas distributions, which could undergo a mixture of multiple processes and are classified as the candidate RPS sample.
The $20\%$ merger fraction (4/20) is consistent with previous studies of $z \approx 2$ clusters~\cite{LiuShuang:2023,Waston:2019}, which is higher than field galaxies.

The degree of morphological disturbance is further quantified by two parameters: the asymmetry of the CO(3-2) distribution relative to the centre of the stellar component ($A$) and the radius of the CO(3-2) disks ($r_{\rm gas}$); $r_{\rm gas}$ is derived as the mean distance of the pixels with half of the peak CO(3-2) flux from the emission peak, twice of which is the approximation of the full-width at half-maximum of the CO distribution.
The two parameters describe the consistency between gas and stellar distribution and the extension of the gas disks, respectively. 

Figure~\ref{fig3} shows the distribution of the 16 non-merging CO(3-2)-detected members in the morphological parameter space. The visually classified disturbed galaxies exhibit systematically larger $A$ and{/or} $r_{\rm gas}/r_{\rm eff,\star}$ ($r_{\rm eff,\star}$ is the effective radius from ALMA point spread function (PSF)-convolved F444W image), while the galaxies classified as `normal' (N1 to N7, CO moment-0 maps shown in Extended Data Figure 3) {and the field galaxies at similar redshift} are more symmetric and compact. 
{Two of the high-fidelity candidates (D2 and D4) have $A$ values that fall within the demarcation boundary. Excluding N4, which hosts a radio active galactic nucleus (AGN)~\cite{Wang+16}, D2 and D4 nevertheless lie towards the higher end of the asymmetry distribution compared to most normal and field galaxies.
 }
 
This verifies that the two populations have distinct molecular gas distributions. 
Given that the typical positional accuracy is better than $0.2''$ (Methods), the asymmetric structures cannot be explained solely by positional uncertainties, and external forces competing with the galaxy's anchoring pressure should come into effect. 
{In Extended Data Figure 4, we also show the significance of the extent structures in some disturbed galaxies.}
Although some normal galaxies exhibit disturbed features in the outskirts (such as N2 and N5), the signal is much weaker than their central high-surface-density regions (the central gas density is over four times higher than that in the outer parts, as shown in Extended Data Figure 3).

In general, two different kinds of environmental effects could account for the observed disturbed gas morphology: tidal interactions from galaxy-galaxy interactions~\cite{Combes:1988,Alberts+2022,Moore+1998,Merritt+1983} and RPS from galaxy-intracluster medium (ICM) interactions~\cite{Boselli+2006,Noble+2019,Boselli+2019,Boselli+2022,Cramer:2020,Gunn+1972}.
The major difference between the two is that tidal stripping would also strip and influence the stellar components.
From a morphological perspective, by examining the concentration and asymmetry of the stellar light distributions, we find that the disturbed galaxies in CLJ1001 have similar stellar morphologies to those of the normal cluster members and field galaxies at the same redshift, except for D4 (panel (a) in Extended Data Figure 5). This provides evidence for the absence of strong tidal interactions and for the necessity of RPS.
From a kinematic perspective, the extended gas structures of some disturbed galaxies (such as D1, D4 and D5 in the lower panel of Fig.~\ref{fig2}) inherit the kinematics from the rotating disk of their hosts, indicating their origins in the host galaxies, as also seen in  simulations~\cite{Kronberger+2008,Boselli+2021} and observations of RPS events in the nearby Universe~\cite{Fumagalli+2014,Sardaneta+2022}. 
The ordered rotation from either double-horn profiles (for example, D1 and D7; Extended Data Figure 6) or moment-1 maps (such as D1 and D4; Extended Data Figure 7) also corroborates the lack of strong tidal forces. In addition, most disturbed galaxies have no close neighbours (Fig.~\ref{fig2}) except for D5, D6 and D7. For D7, the cluster member in the northeast (N5) has a velocity difference of $\sim 1500~\rm km/s$. For D5 and D6, only one close neighbour has spectroscopic redshift (N6) with a velocity difference $>500~\rm km/s$ (the others are H$\rm\alpha$ emitters based on narrowband imaging). However, the tidal radius is 4-5 times larger than the {stellar} effective radius of D5 and D6, indicating that tidal interactions from neighbours are unlikely to play a dominant role in disturbing their gas components. {Similar to cluster galaxies at lower redshifts~\citep{Gavazzi+2001,Vollmer+2003,Cortese+2007}, these galaxies are probably suffering from several perturbing mechanisms, by which tidal interactions could aid RPS via loosening the potential well.}

We further explore the major mechanisms leading to the disturbed gas distribution in cluster galaxies by means of their resolved stellar populations. A recent study~\cite{Lassen+2025} substantiated the different responses of young and older stellar populations to RPS. While the tidal stripping affects stars of all ages similarly, RPS would more efficiently affect the newly formed stars by disturbing the gas from which they are born. Comprehensive resolved spectral energy distribution (SED) fitting is challenging as only four bands of JWST imaging are available; we therefore explore this tendency using the F150W--F277W colour diagram {(F150W and F277W correspond to rest frames of $\sim$4,300 \AA{} and 7,900 \AA{}, respectively)}. We calculate the asymmetry for each cluster galaxy in the F150W-F277W colour diagram and find that the disturbed galaxies are more asymmetric in the spatial colour distribution { but have no obvious differences in the stellar shape asymmetry compared with field galaxies at the same redshift with controlled stellar masses} (Extended Data Figure 8), which further supports the RPS scenario.
Therefore the gas disturbance in most of the disturbed galaxies is unlikely to be caused by tidal stripping or mergers, and RPS should be a more plausible mechanism. The participation of gravitational interactions (such as harassment~\cite{Moore+1998,Duc+2008,Vollmer+2005}) in addition to RPS could not be fully ruled out for disturbed galaxies without typical gas-stripping tails.

In local clusters, there have been observations of molecular gas and dust stripped from the host galaxy by ram pressure~\cite{Jachym+2014,Jachym+2017,Longobardi+2020,Souchereau+2025,Cramer:2020}. At high redshift, RPS has been observed up to $z \approx 0.7$ in ionized gas~\cite{Boselli+2019}, and potentially at $z \approx 1.6$ in the molecular gas~\cite{Noble+2019,Cramer+2023}. For galaxy clusters at even higher redshifts, despite their generally low halo masses, RPS is still expected to take place given their higher ICM densities~\cite{Boselli+2022}. Compared with local clusters, the ICM density in CLJ1001 should be over 1 dex higher due to the higher critical density of the Universe.
The detections of extended X-ray emission~\cite{Wang+16} and Sunyaev–Zeldovich effect (R. Gobat et al., manuscript in preparation) in CLJ1001 also show that hot ICM is already present in this young cluster. Under the RPS scenario, we calculate the truncation radius of the CLJ1001 galaxies, which is around 1--2 effective radii of their gas distribution ({except for D3 and D9}; see Methods). Hence about 20--50\% of the gas could be easily disturbed and stripped from their host galaxies.
CLJ1001 is found to lie at the intersection of several filaments (panel (b) in Extended Data Figure 9). The disturbed galaxies at outskirts locate well on the filaments. These galaxies may directly fall towards the centre of CLJ1001, following the filament's trajectory at high differential velocity, and potentially interact with the dense clumps within the filaments~\cite{Angelinelli+2021}, thereby undergoing more effective RPS.

To gain further insights into the nature of the disturbed galaxies, we examine their distribution in the phase-space diagram in the right panel of Fig.~\ref{fig3}.
The locations of four disturbed galaxies (D5, D6, D7 and D9) in Fig.~\ref{fig3} are generally consistent with RPS galaxies seen in simulations or in local galaxy clusters~\cite{Jaffe:2015,Jaffe:2018}, in the inner cluster regions with a high density of ICM. They probably have been accreted onto the cluster early and have been substantially stripped, as supported by their low gas fractions.
The other disturbed galaxies are located in the outer cluster region with $R \ge 1\,R_{200c}$, but have low relative velocity differences from the cluster centre. This means that these galaxies fall nearly straight onto the cluster centre with high radial velocities {or have large peculiar velocity in the background or foreground of the cluster centre, resulting in small redshift offsets}; this large differential velocity is the dominant factor of the strong RPS event, as also shown in simulations~\cite{Jaffe:2015,Arthur:2019}. Different from the RPS candidates in the centre, those in the outer region are most likely to be going through their first infall, as shown by their relatively high gas contents. That is, the RPS may have just started for many of them as we observed, and has not removed a large portion of their cold gas. 
Moreover, three of the four mergers have small line-of-sight velocities relative to the cluster centre, which could be a promising condition for mergers taking place in the low-velocity-dispersion region of CLJ1001.

\subsection*{Kinematic and star-formation properties of disturbed galaxies}
To explore the effects of gas stripping on the properties of these cluster galaxies, we examine the stability in gas disks by deriving the resolved Toomre {$Q$} parameter from CO(3-2) moment maps (see Methods). We show the cumulative distribution function (CDF) of median {$Q$} within stellar effective radius in Fig.~\ref{fig4}. The disturbed galaxies have larger {$Q$} than the normal galaxies, except for N4. 
N4 hosts a radio AGN, whose gas could be disturbed by the activity of the AGN~\cite{Murthy:2022}. The other normal galaxies mostly show regular rotation with double-horn profiles (Extended Data Figure 6). Excluding this outlier, we plot the mean distribution of {$Q$} along the distance to the galaxy centre in Fig.~\ref{fig4}. Again, the disturbed galaxies exhibit larger {$Q$} within approximately the inner 1.5 $r_{\rm eff,\star}$, than the normal galaxies, suggesting more stable gas disks that contradict their star-forming states. Such high {$Q$} values are seen in simulated galaxies undergoing perturbations~\cite{Inoue:2016}, with gas compression in the turbulence and star-forming clumps forming later. As {$Q$} is proportional to the velocity dispersion and anticorrelated with the gas surface density, it also provides evidence for the stripping scenario, which leads to a low gas surface density and increased velocity dispersion in turbulence against the self-gravitational gas disk.

Molecular gas is the fuel for star formation, and in the right panel of Fig.~\ref{fig4}, we find that the disturbed galaxies exhibit higher star formation efficiencies (SFE$={\rm SFR}/M_{\rm gas}$; where SFR is the star formation rate and $M_{\rm gas}$ is the molecular gas mass), about 0.35 dex higher than the normal galaxies.
This seems to be mainly driven by their systematically lower gas surface densities, which are $\sim$0.56 dex lower than those of normal SFGs in CLJ1001. 
Alternatively, enhanced star formation has also been observed in RPS events
up to $z \approx 0.5$ ~\citep{Vulcani+2018,Vulcani+2024}. These features provide support for an RPS origin for the disturbed galaxies, which globally exhibit low gas densities at given local gas surface densities due to RPS; in tandem, shock compression may increase the fraction of dense gas and hence further contribute to an enhanced SFE.\\\\

Our results have revealed various environmental effects at work in the forming cluster CLJ1001 at redshift $z = 2.51$. 
In particular, this work highlights the occurrence of RPS events in clusters at the highest redshifts.
While tidal interactions are expected to be common in these high-z, compact (proto-)cluster cores, RPS can already operate efficiently only 3 billion years after the Big Bang, implying that a sufficiently dense ICM must have been in place.
This probably reflects the highly stochastic assembly of clusters at early cosmic times, characterized by complex infall trajectories, the convergence and merging of filaments, and—critically—the early emergence of abundant, clumpy ICM in cluster cores, as recently confirmed at z $\approx$ 4~\cite{ZhouDazhi+2025}.
The environmental effects in CLJ1001-like structures in the early Universe that have relatively low halo masses and are actively assembling demonstrate that the quenching of massive cluster galaxies happens not only in mature clusters, but also in an earlier epoch.

\section*{Methods}\label{sec11}

Throughout this work, we adopt the standard $\Lambda$ cold dark matter (where $\Lambda$ is the cosmological constant) cosmology with the Hubble constant at the present day $H_0=70\,\rm km\,s^{-1}{\rm Mpc}^{-1}$, the present-day matter density parameter $\Omega_{\rm M}=0.3$ and the present-day dark energy density parameter $\Omega_{\rm \Lambda}=0.7$, and the stellar initial mass function from ref.~\cite{Kroupa:2002}.

CLJ1001 is located in the COSMOS field with multiband observations, including the Visible and Infrared Survey Telescope for Astronomy (VISTA)/K band~\cite{McCracken:2012}, JWST/NIRCam~\cite{Casey:2023}, Subaru/Multi-Object Infrared Camera and Spectrograph (MORICS) CO narrowband (rest-frame $\approx 2.3\,\rm\mu m$) observations (PI: T. K.), and the JWST COSMOS-Web survey~\cite{Casey:2023}. More details are summarized in ref.~\cite{Sun:2024}.

\subsubsection*{Source detection and classification with ALMA data}

We carried out observations of the CO(3-2) transitions (rest-frame frequency 345.796~GHz) for galaxies in CLJ1001 at $z=2.51$ with ALMA band 3. These observations were taken between 2016 November 21 and 2018 January 13 (Project IDs 2016.1.01155.S \citep{Xiao+22} and 2017.1.01099; PI: Tao Wang). Project 2016.1.01155.S in ALMA Cycle 4 consisted of two different array configurations towards the pointing at the cluster centre, the more compact C40-4 and the more extended C40-7 configurations, with integration times of $\sim$1.1 h and 2.2 h, respectively. Project 2017.1.01099 in ALMA Cycle 5 targets the two upper areas and one lower area out to $\sim 3R_{200\rm c}$, observing in array C43-6 with an integration time of $\sim$ 4.2 h for each pointing. The full-width at half-power of the ALMA primary beam for each pointing is $\sim 53''$.  

We performed the data calibration using pipeline in the Common Astronomy Software Application package~\cite{McMullin:2007} (CASA; in version 4.7.2, 4.7.0 and 5.1.1 for individual observations). 
We also used CASA 6.2.1.7 for data products and visualization after the calibration. We combined the visibility data and used \textsc{tclean} with a 0.2$''$ cell size, a uvtaper of 0.6\,$''$, a channel width of 16~MHz, a Briggs weighting of robust$=$2 and a $2.5~\sigma$ threshold. To obtain the CO(3-2) line data cube, we performed \textsc{imcontsub} task in CASA with linear continuum subtraction. The output data cube has an averaged synthesized beam size of $1.09''\times 0.98''$ (and $1.02''\times0.93''$ for the lower single pointing, $\sim 8~{\rm kpc}$ on the physical scale) and a r.m.s. level of $\sim 90~{\rm\mu Jy\,beam^{-1}}$ per channel at the phase centre.

We also cleaned a high-resolution data cube using the weighted visibility data (for central pointing, we assigned the weights of the compact array to the extended array to be 1:4 to match the average resolution in the other pointings) with a Briggs weighting of robust=1.5 and a uvtaper of 0.3\,$''$ for further data presentation. The resolution of the returned high-resolution data cube is $0.66''\times0.58''$.

Ref.~\cite{Sun:2024} selected a list of cluster members based on the photometric redshifts and H$\rm \alpha$ emission. We first extracted the spectrum from CO(3-2) data cube centred at these H$\alpha$ emitters, which returns 19 detections with a signal-to-noise ratio ($\rm SNR$) $>4$ (except for weak detections of D8 and D9 with SNR $\approx$ 3), including the 4 galaxies presented in ref.~\cite{Xiao+22} and 13 of the CO(1-0)-detected galaxies presented in refs.~\cite{Wang+16,Wang+18}. The velocity range was determined from the velocity-integrated fluxes with maximum SNR. We also detected sources directly from the CO(3-2) data cube. For every pixel in the data cube, we subtracted the spectrum within 2$\times$ the synthesized beam , collapsed the data cube in the velocity range and then flagged it if it was the local brightest peak. If so, we tried to find an optical counterpart with a photometric redshift of $\sim$2.51. We required the source to have SNR $\ge$ 4 within a synthesized beam. The SNR is calculated as \cite{Lenz:1992,Anderson:2023}: 
\begin{equation}
    {\rm SNR}=0.7\left(\frac{\Delta V}{\Delta v}\right)^{0.5}\frac{f_{\rm peak}}{\sigma_{\rm RMS}},
\end{equation}
where $\Delta v$, $f_{\rm peak}$ and $\sigma_{\rm RMS}$ are the spectrum resolution, peak flux and r.m.s. The r.m.s. is estimated in the negative channels. The 0.7 value corresponds to a Gaussian line profile. $\Delta V$ is the full-width at half-maximum (FWHM) of the emission line obtained by fitting the Gaussian to the spectrum. In this way, we gained one new detection (galaxy N1) in addition to the previous 19 detections. We list the information on these sources in Extended Data Table 1. 

With a large sample of spectroscopic members (including an extra four H$\rm\alpha$ emitters from ref.~\cite{Wang+16}), we re-derive the central redshift of the cluster by fitting a Gaussian function to the redshift distribution (using \textsc{curvefit} in \textsc{SciPy}~\cite{scipy:2020}), yielding $z_{\rm c}=2.510\pm0.002$ (Extended Data Figure 9). This is slightly different from the value reported in ref.~\cite{Wang+16} ($z_{\rm c}=2.506$), which is probably due to the smaller sample of spectroscopic members in ref.~\cite{Wang+16} and the fact that the early confirmed members include mostly galaxies in the cluster core.

{ We also identified the cosmic filament with \textsc{DisPerSE}~\citep{Sousbie+2011a,Sousbie+2011b} based on galaxy catalogue from COSMOS-Web~\citep{Shuntov+2025}. For each galaxy we drew Monte Carlo redshift realizations from its photometric-redshift probability distribution function, PDF($z$), generated multiple realizations of the projected galaxy distribution in each slice and extracted a \text{DisPerSE} skeleton from each realization (using a 2$\rm \sigma$ \textsc{-nsig} threshold). We applied the method to a single slice spanning $z$=2.48-2.54, and adopted all galaxies with $\int_{2.48}^{2.54}{\rm PDF}(z){\rm d}z>0.05$ and stellar mass $M_\star>10^8~{M_\odot}$. We included the  narrowband-selected $\rm H\alpha$ emitters~\citep{Sun:2024} with redshifts fixed at the cluster centre and utilized all galaxies with available spectroscopic redshifts ~\citep{Wang+16,Wang+18,Sun:2024}. We then stacked the resulting skeletons into a filament probability map ($w_{\rm fil}$) on a 250 comoving kiloparsec (ckpc) grid by summing, in each cell, the lengths of filament segments weighted by their \textsc{DisPerSE} robustness, and re-distributed $w_{\rm fil}$ with a cloud-in-cell scheme to enhance continuity. Finally, we re-extracted the filament network from the $w_{\rm fil}$ map using the optimized persistence \textsc{-cut} thresholds~(Q.H. et al., manuscript in preparation). The result is shown in the right panel of Extended Data Figure 9. CLJ1001 is at the interface of several filaments on a large scale and the RPS candidates locate well on these identified filamentary structures. Compared with galaxies in the CLJ1001 core, those disturbed galaxies along the outer filament spine are more massive than the surrounding ones and inhabit a lower-density environment, where strong tidal interactions are unlikely and RPS is expected to be more effective along filaments~\citep{Bahe+2013}.}

We used task \textsc{immoments} in \textsc{CASA} to create the CO(3-2) moment-0 (flux), moment-1 (line-of-sight velocity) and moment-2 (velocity dispersion) images. We corrected the effect of beam smearing in moment-2 maps by deducting the standard deviation of moment-1 maps within a synthesized beam at each pixel. As low-SNR pixels in moment-0 could result in unreliable velocity subtraction, we used the software package \textsc{VorBin}~\cite{Cappellari:2009} to make bins of SNR$>$5 based on the moment-0 maps and then binned the moment-1, moment-2 maps in the same manners. We show the moment-1 images in Extended Data Figure 7.

We fitted the S$\rm\acute{e}$rsic profile using \textsc{PetroFit}~\cite{Geda:2022} for cluster galaxies in F444W images, which returns the effective radius ($r_{\rm eff,\star}$), S$\rm\acute{e}$rsic index, position angle and centre position of the stellar component.

As the ALMA observation has a larger PSF than the JWST observation, we should only use the effective radius from the F444W image after convolution with the ALMA synthesized beam when comparing with the ALMA data. We first convolved the F444W image with the ALMA synthesized beam and then fitted it with a Gaussian using \textsc{PetroFit} at the centre from the fitting in the unconvolved F444W image. We adopted the half-light radius of the output Gaussian model as the convolved effective radius (which is also the FWHM for a 2D Gaussian). We note that we mean the convolved effective radius when comparing between the ALMA and JWST observation.

In Extended Data Figure 5, we also show the distribution of the stellar concentration and asymmetry of cluster members and some field galaxies in the COSMOS field using unconvolved F444W imaging at $2<z<3$ computed from \textsc{GaLight}~\cite{Ding:2021} (images for field galaxies from ref.~\cite{Sun+2026}).

As the cluster galaxies undergoing environmental effects could have asymmetric or one-sided gas distributions, to classify galaxies on the basis of their gas morphologies, we defined an approximate radius $r_{\rm gas}$, for the gas disks. To calculate $r_{\rm gas}$, in CO(3-2) moment-0 maps for each galaxy we reserved only pixel values with SNR $\geq \sqrt{2}\,\sigma$, where $\sigma$ is the r.m.s. of the negative values in the map. 
We also tested with the 1$\sigma$ and 2$\sigma$ thresholds and the results are consistent.
We used the function \textsc{removing\_small\_objects} in Python package Scikit-image~\cite{Walt:2014} to reserve the largest connected domain associated to the host galaxies to avoid noise. The peak of the CO(3-2) emission was determined using Python package \textsc{Photutils}~\cite{Bradley:2020}; $r_{\rm gas}$ was then derived from the mean distance of pixels with half of the flux of the maximum from the peak position. 

Next, we derived the non-parametric morphology parameter, asymmetry ($A$), following refs. \cite{Conselice:2003,Nurgaliev:2013}:
 \begin{equation}
     A=\frac{\sum|I_0-I_{180}|}{2\sum|I_0|},\label{equ2}
 \end{equation}
where $I_0$ and $I_{180}$ are the original image and image rotated 180 degrees around the centre position of the stellar component. To estimate the errors of $r_{\rm gas}$ and $A$, we performed 1,000 random runs. In each run, we calculated the two parameters for a noise-added Gaussian model with the same CO(3-2) peak amplitudes and stellar effective radii as the cluster galaxies, and adopted the standard deviation as the errors.

We plot the galaxies in the $r_{\rm gas}/r_{\rm eff,\star}-A$ plane in the left panel of Fig.~\ref{fig3}. The clear existence of a group with larger $A$ and{/or} larger $r_{\rm gas}$ over the stellar component indicates asymmetric elongated gas structures. Thus we used the demarcation line shown in Fig.~\ref{fig3} to divide disturbed and normal galaxies. We also tried to fit a Gaussian to the CO(3-2) distribution, and the result is consistent with that shown in the {right} panel of Extended Data Figure 5. But because of the asymmetric gas morphology, a single Gaussian could not fully describe their distribution. The merger and disturbed galaxies are shown in Fig.~\ref{fig2}, and the normal galaxies are shown in Extended Data Figure 3.

For better presentation, we also contoured the moment-0 maps integrated in the same velocity range from the high-resolution data cube on the JWST RGB image in Extended Data Figure 1. The disturbed galaxies could also be identified with the off-centre/elongated features from the high-resolution image.

To further show the significance of the emission outside the main group of galaxies, we compared these signals to the background noise level. First, in the moment-0 maps we fitted the Gaussian function to the distribution of noise within a synthesized beam exclusively with negative values to avoid including any unclassified emission. The fluxes from apertures of the same shape as the synthesized beam around the galaxies were then derived and compared to the noise standard deviation. We show the significance of the extended emission for D1-D6 in Extended Data Figure 4.

We note that the above significance test would underestimate the significance of our disturbed galaxies, because the continuity in the velocity, the irregular and extended shapes in spatial distribution have not been fully accounted for.

To further verify the disturbance in these galaxies originates from the dense environment or from the intrinsically irregular shape at high redshift, we used a sample of field galaxies with JWST F444W imaging (JWST images from refs.~\cite{Sun:2024,Sun+2026,Casey:2023,D'Eugenio:2024}) and ALMA CO(3-2)/CO(2-1) observations (D.L. et al., manuscript in preparation; CO data were collected from ALMA archive with the following programme IDs: 2021.1.01291.S, 2017.1.00270.S, 2018.1.00543.S and 2015.1.01590.S). After decreasing the resolution to $\sim 1.1''$, we performed the same morphological diagnostics used for the cluster galaxies. The result is shown in {Fig.~\ref{fig3}}. Most field galaxies locate in the same region as the normal galaxies in CLJ1001, which cannot explain the large fraction of disturbed galaxies in CLJ1001.

\subsubsection*{Derivation of physical properties of the cluster galaxies}
To examine whether the young and old stellar populations react similarly to gas stripping, we calculated the asymmetry in the F150W-F277W colour maps for each cluster galaxy (using the Equation~\ref{equ2}). 
We then plotted the CDF of this colour asymmetry for different categories of galaxies. In Extended Data Figure 8, we show that the disturbed galaxies generally have larger asymmetry in the colour maps. We also derive the colour asymmetry for field star-forming galaxies in COSMOS within redshift bin of {[2.26,2.76]} (catalogue {and images} from ref.~\cite{Sun+2026}) for comparison after controlling for the stellar mass with the disturbed galaxies, and found that the disturbed galaxies are more asymmetric than the field galaxies. This difference suggests that the stars of different ages have different responses to gas stripping, or the dust has been redistributed, which are both the positive evidence for RPS influencing the young stars or dust extinction~\cite{Abramson+2011,Abramson+2016,Lassen+2025}.
{We also performed a comparison of the stellar shape asymmetry from the F444W imaging between the disturbed galaxies and stellar-mass-matched field star-forming galaxies (using Equation~\ref{equ2}, where $I_0$ are the segmentation maps). The edge detection was set to 8$\sigma$ above the background r.m.s. N6 was not included because it could not be deblended from the two nearby galaxies. The disturbed galaxies have a similar distribution to the  field galaxies, suggesting the non-dominant role of tidal interactions. The normal galaxies have smaller asymmetry, mainly due to the larger stellar mass.}

To derive the masses of molecular gas ($M_{\rm gas}$) and the gas fractions ($\mu_{\rm gas}{=M_{\rm gas}/M_\star}$), in CO(3-2) moment-0 image, we first obtained the resolved velocity integrated flux ($S_{\rm CO}\Delta v$). By using equation 3 in ref.~\cite{Solomon:2005} to convert $S_{\rm CO}\Delta v$ into CO(3-2) line luminosity ($L^{\prime}_{\rm CO\,3-2}$), the line ratio between CO(3-2) and CO(1-0), $R_{31}=0.7$ (mean value in ref.~\cite{Xiao+22}), the $\rm CO$-to-$\rm H_2$ conversion factor $\alpha_{\rm CO}=4.10$ (mean value in ref.~\cite{Wang+18}) and a factor of 1.36 for the helium contribution, we derived the resolved $M_{\rm gas}$ for these galaxies. The $\mu_{\rm gas}$ values were determined from $\mu_{\rm gas}$ within the ALMA synthesized beam-convolved stellar effective radius (where the stellar mass is approximated to be the half of the total); {$\mu_{\rm gas,MS}$ was derived using a scaling relation from ref.~\cite{Tacconi:2018}.}

\begin{figure}
\centering
\includegraphics[width=1\textwidth]{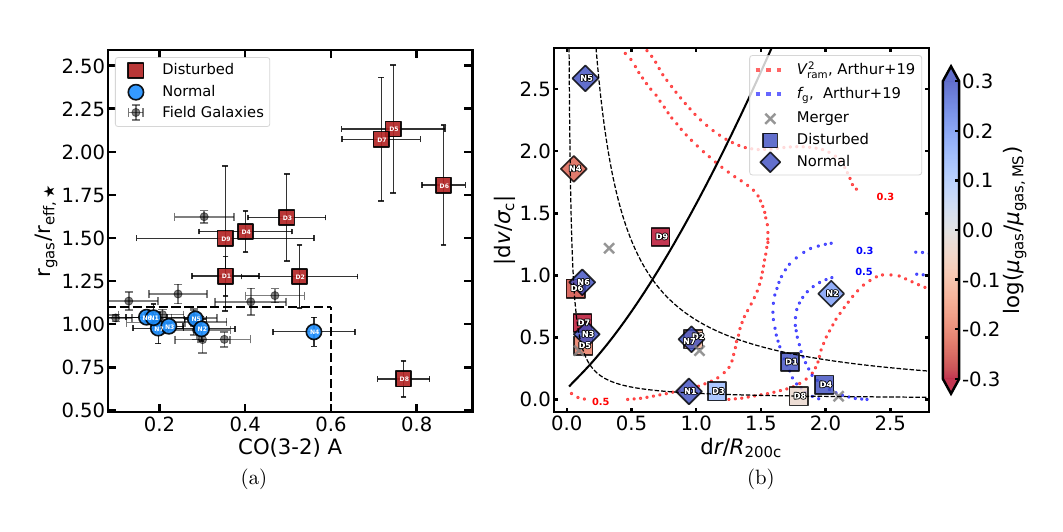}
\caption{\textbf{The morphological diagnostics and phase-space diagram of member galaxies in CLJ1001.} 
(a) Plot of $r_{\rm gas}/r_{\rm eff,\star}$ versus $A$. The dashed line shows the demarcation between disturbed and normal galaxies. Error bars show 1$\sigma$. (b), Phase-space plot for the galaxies in CLJ1001. The $x$ axis is the projected radius in sky (d$r$) normalized by $R_{\rm 200c}$ and the $y$ axis shows the line-of-sight velocity offset from the cluster centre (d$v$) normalized by the cluster velocity dispersion ($\sigma_{\rm c}$). $ {\rm d}v=| z_{\rm i}-z_{\rm c}|/(1+z_{\rm c})\times c$, where $c$, $z_{\rm i}$ and $z_{\rm c}$ are the velocity of light and the redshifts of galaxies and CLJ1001, respectively. The two black dashed lines denote $ |{\rm d}v/\sigma_{\rm c}|\times({\rm d}r/R_{\rm 200c})=$0.05 and 0.64, where $\sigma_{\rm c}$ and $R_{\rm 200c}$ are from ref.~\cite{Wang+18}. The blue and red dotted lines from ref.~\cite{Arthur:2019} show the normalized gas fraction ($f_{\rm g}$) of infalling subhaloes and differential velocities ($V_{\rm ram}$) between infalling subhaloes and the ICM at different positions in the phase-space diagram, with the inline values showing the fractions of peak values. The dotted lines show that the increasing differential velocity is the dominant reason for the stronger RPS at the outskirt of the cluster. The black solid line shows the required differential velocity to truncate the gas disk at $2\times$ the gas effective radius for a $\log(M_\star/M_\odot)=10$ galaxy ($\sim$6kpc) by RPS. The galaxies are colour coded by the ratio of their gas fraction ($\mu_{\rm gas}$) to that of the field main sequence galaxies ($\mu_{\rm gas,MS}$).}\label{fig3}
\end{figure}

To derive the SFR and $M_\star$, we performed SED fitting with BAGPIPES~\cite{Carnall:2018}. We applied the delayed star-formation history, the Calzetti extinction law~\cite{Calzetti:2000}, a stellar
population synthesis model from the 2016 version of BC03~\cite{Bruzual:2003} and nebular emission~\cite{Byler:2017}; more details are summarized in ref.~\cite{Sun:2024}. We show their locations relative to the star-formation main sequence in Extended Data Figure 2.
Star formation is enhanced for D1 and D4, possibly due to RPS~\cite{Vulcani+2024}. The SFRs for the remaining disturbed galaxies are marginally below the main sequence, and their gas fractions are reduced (Fig.~\ref{fig3}), which is evidence for the advanced phase of gas stripping.
The SFR in the effective radius is simply assumed to be the half of the total SFR.

To derive the truncation radius ($R_{\rm trunc}$), we used the classical criterion for ram-pressure stripping~\cite{Gunn+1972}:
\begin{equation}
    \rho_{\rm ICM}\,V^2\ge2\pi\,G\,\Sigma_{\rm \star}\,\Sigma_{\rm gas}.\label{eq3}
\end{equation}
We adopted the ICM models from \cite{Patej:2015,Morandi+2015,Boselli+2022}, using equation (11) in ref.~\cite{Boselli+2022}:
\begin{align}
[\Omega_{\rm M}(1+z)^3+\Omega_\Lambda]^{-1}\, n_{\rm e}(x)
&= 0.00577
  \left(\frac{x}{0.201}\right)^{-0.150}
  \left(\frac{x}{0.265}\right)^{-0.0638} \nonumber \\
&\quad\times \left[1+0.759\left(\frac{x}{0.201}\right)^{0.949}\right]^{-2.936}
\end{align}
, where $n_{\rm e}$ is the electron density and {$x$} is $\frac{r}{R_{200\rm c}}$
. We adopted cluster mass $\log(M_{\rm 200c}/M_\odot)=13.78$, $R_{\rm 200c}=340\,{\rm kpc}$~\cite{Wang+16,Wang+18}, a baryon fraction of 0.158 and a gas fraction of 0.9.
The ICM density in the centre of the cluster is $\rho_{\rm ICM}=1.81\times10^{-22}{\rm kg/m^3}${ (at 0.02 $R_{\rm 200c}$)}.
We assumed the cluster-centric distance to be $\sqrt{3/2}~\times$ the sky projected distance. We computed the stellar and gas surface density as~\cite{Jaffe:2015,Domainko+2006}:
\begin{equation}
    \begin{aligned}
        \Sigma_\star&=\Sigma_{0,\star}\,e^{-r/(r_{\rm eff,\star}/1.678)},\;
        &\Sigma_{0,\star}=\frac{M_\star}{2\pi\,(r_{\rm eff,\star}/1.678)^2},\\
        \Sigma_{\rm gas}&=\Sigma_{0,\rm gas}\,e^{-r/(r_{\rm eff,gas}/1.678)},\;
        &\Sigma_{0,\rm gas}=\frac{M_{\rm gas}}{2\pi\,(r_{\rm eff,gas}/1.678)^2},
    \end{aligned}
\end{equation}
where the value 1.678 is the conversion factor between the scale length and the effective radius of an exponential disk.

We used the relation in ref.~\cite{Tacconi:2018} to calculate $M_{\rm gas}$ in main sequence galaxies at given stellar masses. {T}he effective radius of gas disks was assumed to be $r_{\rm eff,gas}={1.5}\times r_{\rm eff,\star}$
, because {some recent work revealed extended molecular gas in high-redshift galaxies (especially in (proto-)clusters~\citep{Chen:2024,Rybak+2025}), and the size measurements from F444W imaging are $\sim 20\%$ smaller than those from rest-frame 5000~\AA\; imaging~\citep{vanderWel:2014,Ormerod+2024}}.
Taking an infalling velocity of $1500\,{\rm km/s}$~\cite{Jachym+2017}, comparable to the free-fall velocity from infinity to $1\,R_{\rm 200c}$, we computed the value of $R_{\rm trunc}$ at which Equation~\ref{eq3} would reach the balance. Generally, the cluster galaxies have $R_{\rm trunc} \approx 1$--2 $r_{\rm eff,gas}$, and 20--50\% of the gas would be influenced by the cluster environment{, except for D3 and D9 which are compact and require more clumpy ICM}. 

The RPS criterion (Equation~\ref{eq3}) is applicable to the case where the velocity vector of the galaxy is perpendicular to its gas disk, where RPS is maximized. 
The larger the angle between the velocity vector and the disk normal is, the more difficult it is for the ram-pressure effect to occur. Thus, RPS is a threshold phenomenon. As such, it is likely that the most visible ram-pressure effects occur in the maximum scenario. 
{As the disk angular momentum is conserved, this angle changes along the infalling orbit of cluster galaxies and would reach the efficient face-on case at some point.}
An important consequence in this scenario is that the dislodged gas still in the vicinity of its former host galaxy is expected to display a rotational signature conformal to the host galaxy. Through a closer examination, we found that galaxies D1, D4 and D5 show such signatures. We defer detailed quantification of this signature to future study with the aid of numerical simulations.

To derive the solid line in the right panel of Fig.~\ref{fig3}, we use the mass--size relation from ref.~\cite{vanderWel:2014} { applying a 20\% correction for consistency with size measurement in the F444W band. At each radius, the solid line indicates the required differential velocity for gas truncation at $2~r_{\rm eff,gas}$ in a $\log(M_\star/{M_\odot})=10$ galaxy.}

To derive the Kennicutt-Schmidt law in Fig.~\ref{fig3}, we performed aperture photometry with \textsc{Photutils} within stellar effective radius (convolved with the synthesized beam) on the CO(3-2) moment-0 maps. 
As the resolved SFR is unavailable due to the limited ground-based photometry, we simply assumed half of the total SFR in the effective radius.

To convert the ${\rm SFR}-L_{\rm CO\,3-2}$ relation in ref.~\cite{Tacconi+2013} to the ${\Sigma_{\rm SFR}}-\Sigma_{\rm gas}$ relation, we adopted the same conversion factors as our CO detected galaxies.

\begin{figure}
\centering
\includegraphics[width=1\textwidth]{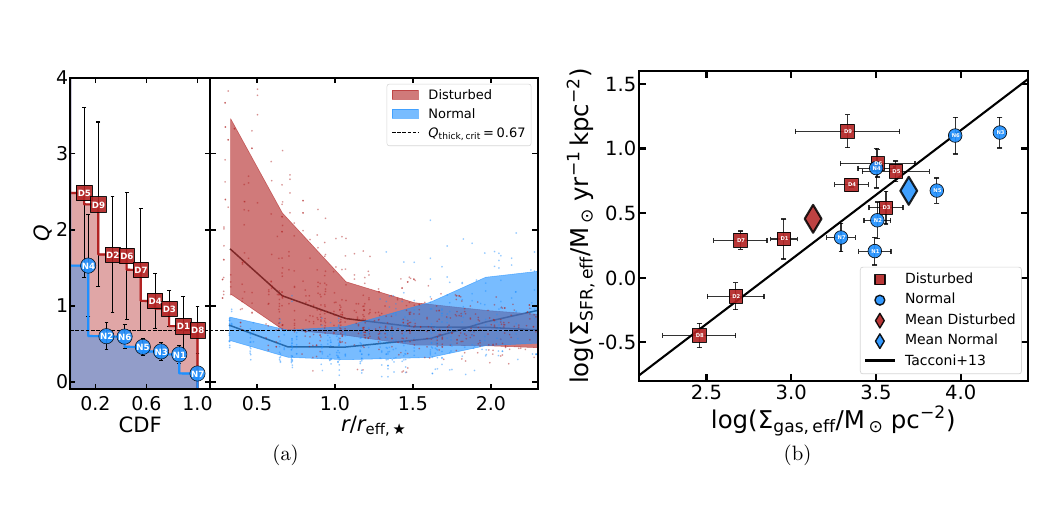}
\caption{\textbf{The Toomre {$Q$} parameter and star-formation relation for cluster member galaxies.} {{(a):}} Left: CDF of the median Toomre {$Q$} parameter within the stellar effective radius of cluster galaxies. The error bars show the 1$\sigma$ standard deviation. The galaxies are shown at their appearance on the CDF. The color-shaded regions is overlain for better display of the CDF. Right: spatially resolved Toomre {$Q$} parameter along the distance from galaxy centres. The small red and blue small dots show the pixel values for disturbed and normal galaxies, respectively, with the solid line and shaded envelopes denoting the median and 1$\sigma$ of the distribution. The horizontal dashed line is the critical $Q$ value for a stable thick gas disk {($Q_{\rm thick,crit}$, see Methods)}, above which the gas disk should be stable. {{(b):}} Kennicutt-Schmidt relation between star-formation rate surface density ($\Sigma_{\rm SFR,eff}$) and molecular gas surface density ($\Sigma_{\rm gas,eff}$) within the stellar effective radius. The solid line is {the star-formation relation for galaxies at $z \approx 1$--3} (see Methods). The error bars show 1$\sigma$. }\label{fig4}
\end{figure}

To derive the spatially resolved {$Q$} parameter~\cite{Toomre:1964} from CO(3-2) data in each pixel, we calculated it following ref.~\cite{Tadaki:2018} as:
\begin{equation}
    Q=\kappa\sigma_{\rm gas}/(\pi G\Sigma_{\rm gas}),
\end{equation}
where $\sigma_{\rm gas}$ is velocity dispersion from the CO(3-2) moment-2 maps corrected for beam smearing, $\Sigma_{\rm gas}$ is gas surface density from the CO(3-2) moment-0 maps and $\kappa$ is the epicyclic frequency. We followed refs.~\cite{Tadaki:2018,Binney:2008} with $\kappa=1.4 V_{\rm max}/r$ at distance $r$ from the galaxy centre; $V_{\rm max}$ was taken as the maximum of $V_{90}-V_{\rm cen}$ and $V_{\rm cen}-V_{10}$, where $V_{90}$, $V_{10}$ and $V_{\rm cen}$ are 90th and 10th percentiles in the CO(3-2) moment-1 maps and the velocity at the stellar centre after inclination corrections. The inclination is determined from the galaxy axis ratio ($b/a$) by assuming the typical intrinsic axis ratio of edge-on galaxies $q_0=0.4$~\cite{Cortese:2016}:
\begin{equation}
    \cos{i}=\sqrt{\frac{(b/a)^2-q_0^2}{1-q_0^2}}.
\end{equation}
{The critical $Q$ value for a stable thick gas disk in Fig.~\ref{fig4} is from refs.~\cite{Genzel:2011,Cacciato:2012}.}

\noindent The position accuracy of the ALMA observations was calculated according to ALMA technical handbook

\begin{equation}
    {\rm pos}_{\rm acc}={\rm beam_{FWHM}}/({\rm SNR}\times0.9),
\end{equation}
where beam$_{\rm FWHM}$ is the FWHM of the synthesized beam and SNR is the signal-to-noise ratio of the image target peak. With a typical SNR of $>$6, the position accuracy is greater than $\sim 0.2''$. For galaxies that are poorly resolved in the low-resolution data cube, the SNR is low because the synthesized beam encompasses a larger area than the galaxy, integrating substantial noise from the surrounding regions. For these galaxies, we examine their SNR ($>$4) in the high-resolution data cube, and they still have the pointing accuracies better than $\sim 0.2''$.

\bmhead{Acknowledgements}

T.W. acknowledges the support of the National Natural Science Foundation of China (grant nos 12525302 and 12141301), the Basic Research Program of Jiangsu (grant no. BK20250001), the National Key R\&D Program of China (grant no. 2023YFA1605600), a Scientific Research Innovation Capability Support Project for Young Faculty (project no. ZYGXQNJSKYCXNLZCXM-P3), the Fundamental Research Funds for the Central Universities (grant no. KG202502) and the China Manned Space Program (grant no. CMS-CSST-2025-A04). Z.-Y.Z. acknowledges the support of the National Natural Science Foundation of China (NSFC) under grant nos 12533003 and 1257030642. R.G. acknowledges funding from project FONDECYT 1231661. D.L. acknowledges the support from the Strategic Priority Research Program of the Chinese Academy of Sciences, grant no. XDB0800401. This work is based (in part) on observations made with the NASA/ESA/CSA JWST. The data were obtained from the Mikulski Archive for Space Telescopes at the Space Telescope Science Institute, which is operated by the Association of Universities for Research in Astronomy, Inc., under NASA contract NAS 5-03127 for JWST. We acknowledge the teams of JWST programmes 1727, 1837, 1840, 2321, 2514 and 3990. We acknowledge the teams of JWST programmes 1895, 1963, 2079, 2514, 3215, 3577, 3990, 6434 and 6541 for developing their observing programme with a zero-exclusive-access period. This paper makes use of ALMA data (project numbers are given in the data availability statement). ALMA is a partnership of ESO (representing its member states), NSF (USA) and NINS (Japan), together with the NRC (Canada), NSTC and ASIAA (Taiwan) and KASI (Republic of Korea), in cooperation with the Republic of Chile. The Joint ALMA Observatory is operated by ESO, AUI/NRAO and NAOJ.

\bmhead{Data Availability}
The ALMA data are available via the ALMA Archive (\href{https://almascience.eso.org/aq}{https://almascience.eso.org/aq}) under project numbers 2016.1.01155.S,
2017.1.01099, 2021.1.01291.S, 2017.1.00270.S, 2018.1.00543.S and 2015.1.01590. The JWST/NIRCam images are available from the
following MAST repositories: \href{https://doi.org/10.17909/1r9y-dv80}{https://doi.org/10.17909/1r9y-dv80}, \href{https://doi.org/10.17909/e1wh-b970}{https://doi.org/10.17909/e1wh-b970} and \href{https://
doi.org/10.17909/8tdj-8n28}{https://doi.org/10.17909/8tdj-8n28}. These data are associated with GO programme 1727 (PIs: J. Kartaltepe and C. Casey), programmes 1837, 1840, 1895, 1963, 2079, 2321, 2514, 3215, 3577 and 3990. 
\bmhead{Code Availability}
All codes used in the paper are publicly available, including \textsc{NumPy}~\cite{numpy}, 
\textsc{Astropy}~\cite{astropy}, 
\textsc{Matplotlib}~\cite{matplotlib}, 
\textsc{PetroFit}~\cite{Geda:2022}, 
\textsc{CASA}~\cite{McMullin:2007},
\textsc{GaLight}~\cite{Ding:2021},
\textsc{BAGPIPES}~\cite{Carnall:2018},
\textsc{Photutils}~\cite{Bradley:2020},
\textsc{SciPy}~\cite{scipy:2020},
\textsc{scikit-image}~\cite{Walt:2014},
\textsc{VorBin}~\cite{Cappellari:2009}
\textsc{DisPerSE}~\cite{Sousbie+2011a,Sousbie+2011b}.

\newpage


\bmhead{Author Contributions Statement}
K.X. reduced the ALMA data, contributed to the main results and wrote the text under the supervision of T.W.
T.~W. initiated the study, interpreted the main results, and improved the text. 
H.S. reduced the JWST data and aided with the photometric measurements for SED fitting. 
L.~C. helped with the galaxy morphology measurements.
Q.~H. helped with constraining the large-scale structure.
E.D., D.E., R.G., A.Z., D.L., M.X., R.C., T.K., C.X., K.K., T.Y., Z.-Y.Z., L.Z. and F.V. contributed to the overall interpretation of the results and the analysis.\\
\bmhead{Competing Interests Statement}
{The authors declare no competing interests.}



\FloatBarrier
\section{Extended Data}
\begin{extfigure}
\centering
\includegraphics[width=0.95\textwidth]{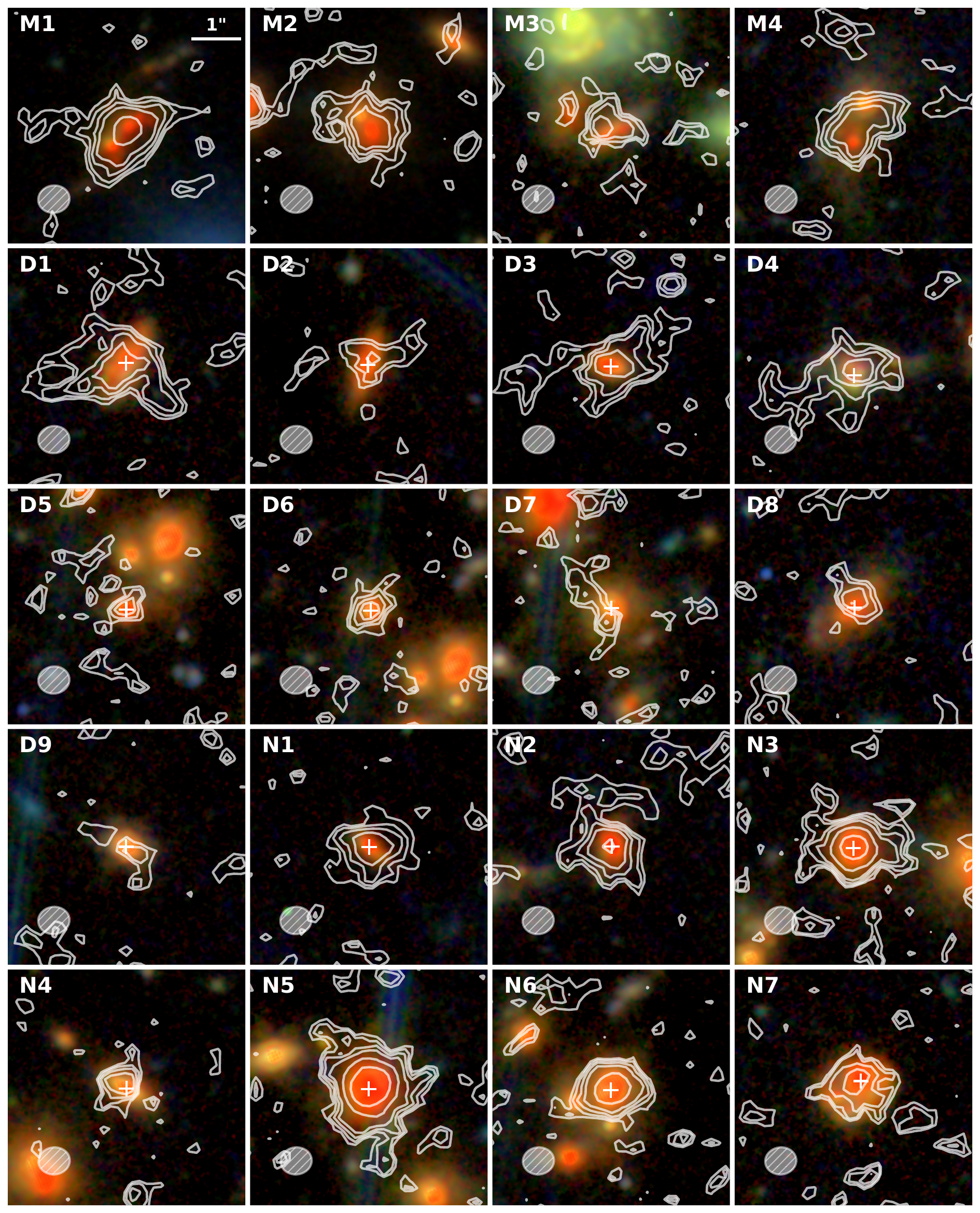}
\caption{\textbf{The high-resolution CO(3-2) distribution for detected cluster members.} The white contours show the CO(3-2) distribution from high-resolution data cube of the member galaxies in CLJ1001 and the levels are $\sqrt{2}$, 2, 2$\sqrt{2}$, 4, 8, 16 times the r.m.s. The false colour NIRCam images are made with F444W (R), F277W (G), and F150W (B) images. All panels are in size of $5''\times5''$. The scale bar is shown in the upper right of the first panel. The lower-left ellipse in each panel shows the synthesized beam of the CO observation. The white plus shows the centre of the galaxy.}\label{efigure1}
\end{extfigure}

\begin{extfigure}
\centering
\includegraphics[width=0.9\textwidth]{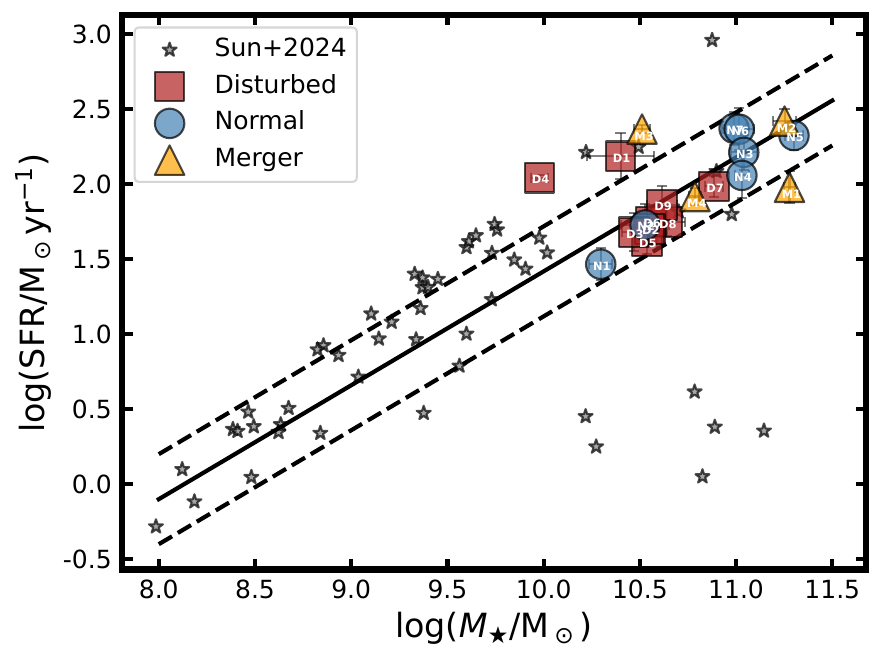}
\caption{\textbf{The plot between SFR and $M_\star$ for cluster members.} The blue circles, red squares and yellow triangles show the normal, the disturbed, and the merger galaxies respectively. The star-formation main sequence is adopted from refs.~\cite{Tacconi:2018,Speagle+2014}, and the dashed lines denote the 0.3 dex offset from the main sequence. The background grey stars are the cluster member galaxies from ref.~\cite{Sun:2024}. The error bars show 1$\sigma$ standard deviation.}\label{efigure2}
\end{extfigure}

\begin{extfigure}
\centering
\includegraphics[width=0.95\textwidth]{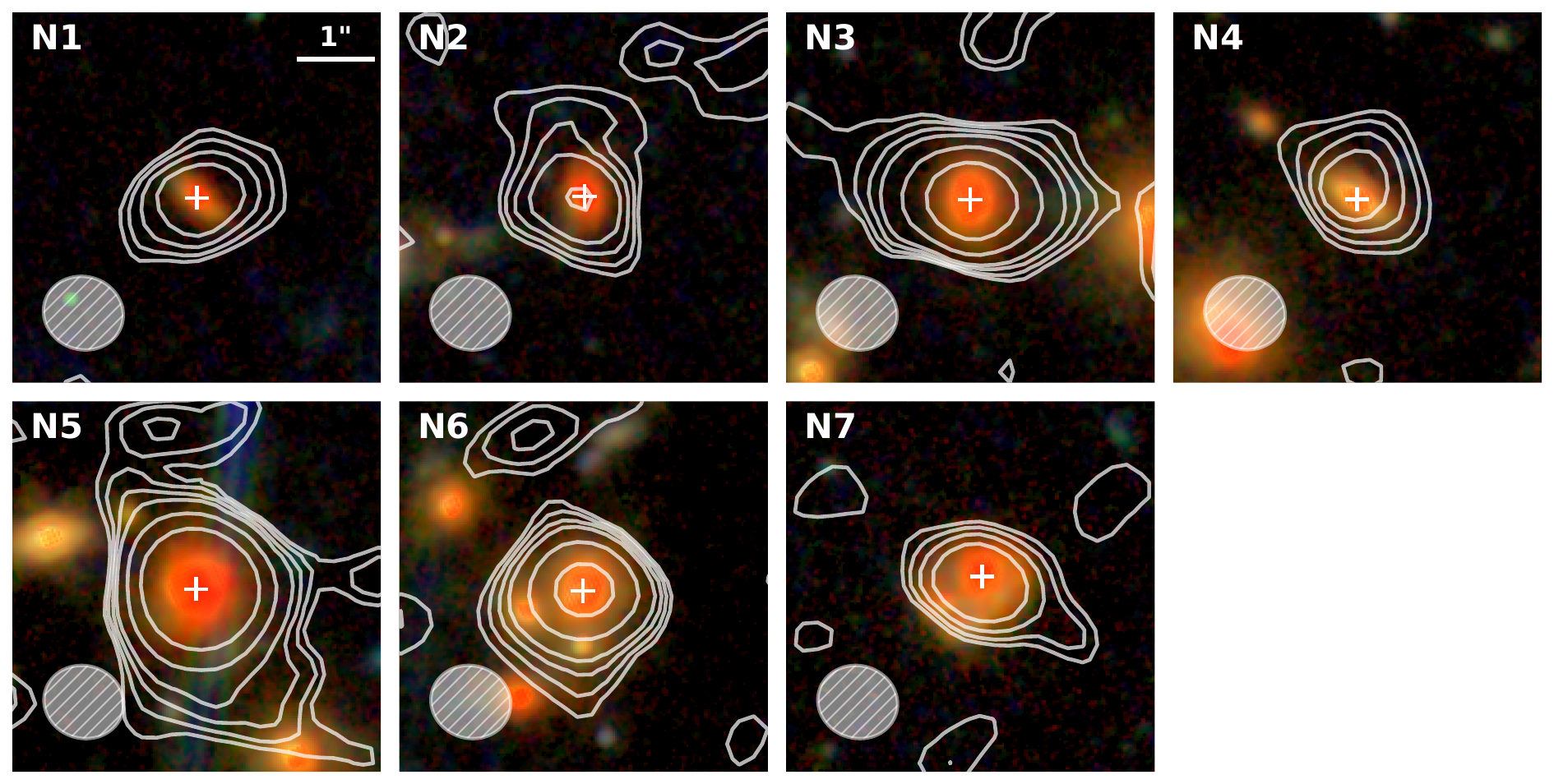}
\caption{\textbf{The distribution of CO(3-2) line emission for the normal cluster galaxies.} The white contours show the CO(3-2) distribution of the normal galaxies in CLJ1001 and the levels are $\sqrt{2}$, 2, 2$\sqrt{2}$, 4, 8, 16 times r.m.s. The false colour NIRCam images are made with F444W (R), F277W (G), and F150W (B) images. The lower-left ellipse in each panel shows the synthesized beam of the CO observation. The white plus shows the centre of the galaxy. All panels are in size of $5''\times5''$. The scale bar is shown in the upper right of the first panel.}\label{efigure3}
\end{extfigure}

\begin{extfigure}
\centering
\includegraphics[width=0.75\textwidth]{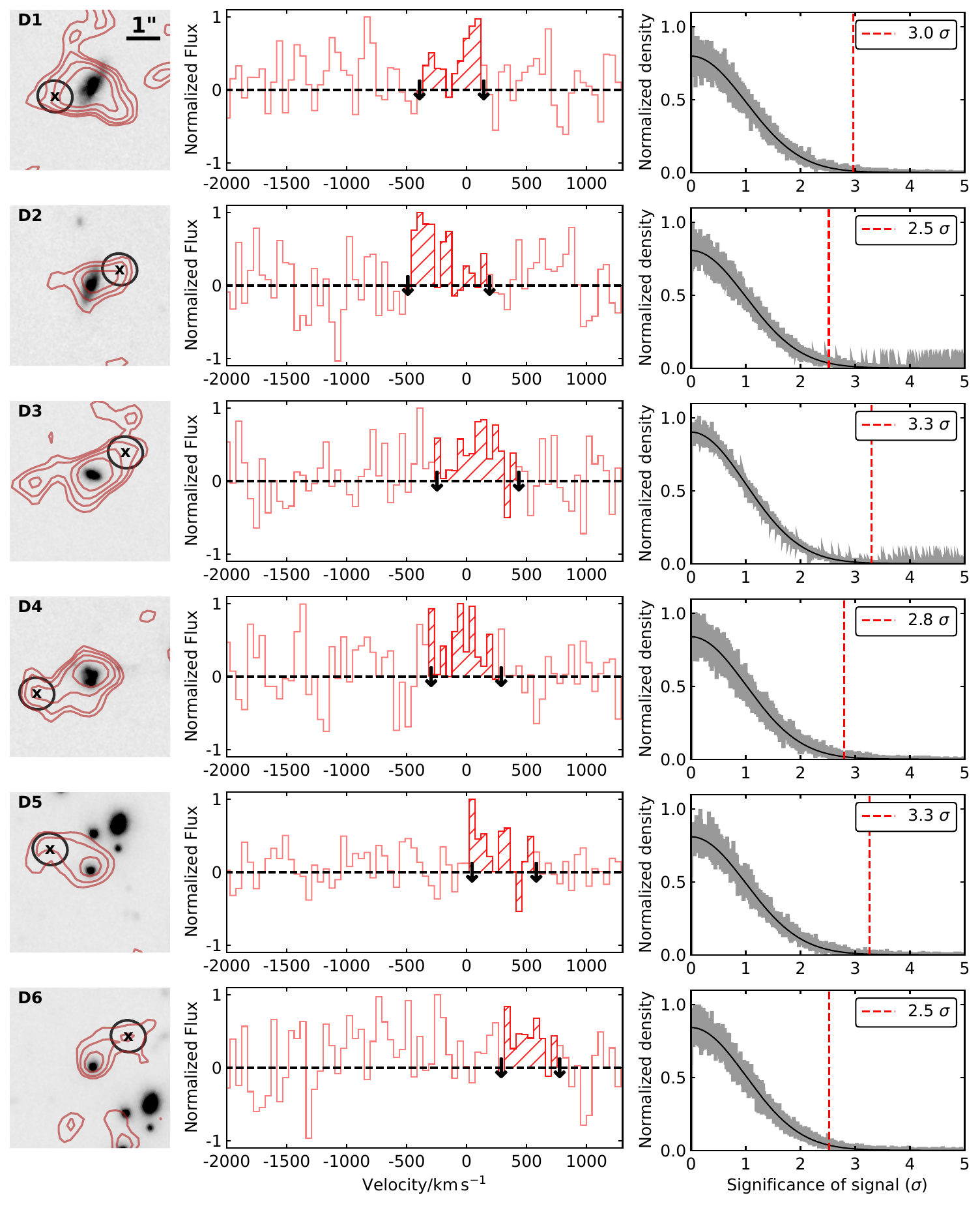}
\caption{\textbf{The significance of the detections of extended structures in the disturbed galaxies.} From left to right: \textit{Left}: The CO(3-2) moment-0 maps (red) contoured on the JWST F444W image. The contours level as [$\sqrt{2}$, 2, 2$\sqrt{2}$, 4, 8, 16] times the r.m.s. The ellipses centred at black crosses denote the interested extended structures. The ellipses are in the same shape as the synthesized beam of ALMA observations. The scale bar is shown in the upper right of the first panel. The panels are in size of $5''\times5''$. \textit{Middle}: The mean spectrum from the ellipses in the left panel. The arrows show the line widths in the host galaxies. \textit{Right}: The significance of detections within the ellipses marked in vertical dashed lines. The grey histogram is the noise distribution from multiple apertures placed at different sky regions and the black solid lines are the best-fitted Gaussian model.}\label{efigure4}
\end{extfigure}

\begin{extfigure}
\centering
\includegraphics[width=1\textwidth]{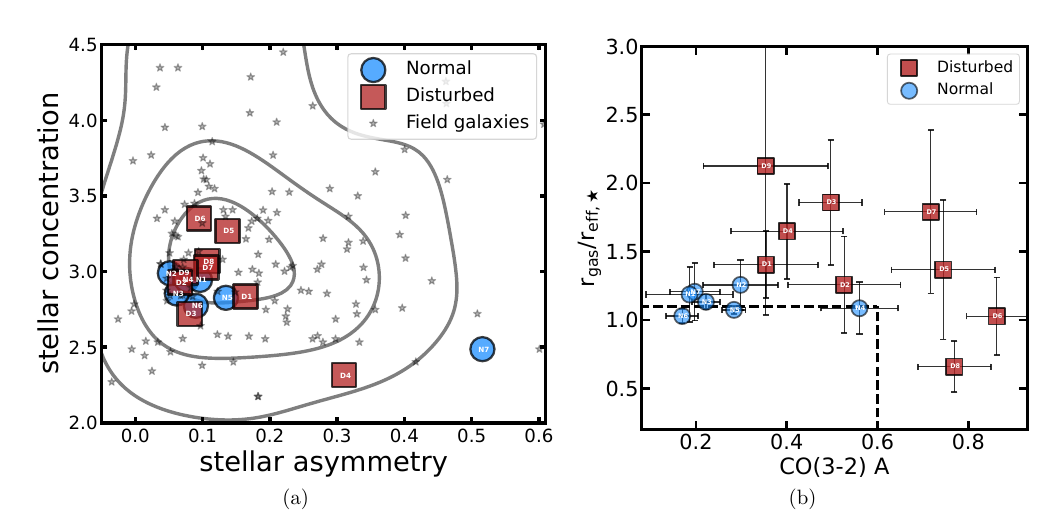}
\caption{\textbf{The concentration vs. asymmetry plane for cluster members and the morphological diagnostics result with \textsc{IMFIT}.} (a) shows the plot between the stellar concentration and asymmetry for normal (blue circles) and disturbed (red squares) galaxies in CLJ1001. The background grey stars are field galaxies at the same redshift and the corresponding contours show their cumulative distribution at 16\%, 50\% and 84\% levels. (b) shows the ratio of $r_{\rm gas}/r_{\rm eff,\star}$ and CO(3-2) asymmetry using the gas size measurement from Gaussian modelling with \textsc{IMFIT}. The error bars show the $1\sigma$ uncertainties. The dashed lines show the demarcation between the disturbed galaxies and normal galaxies adopted in the main text.}\label{efigure5}
\end{extfigure}

\begin{extfigure}
\centering
\includegraphics[width=0.95\textwidth]{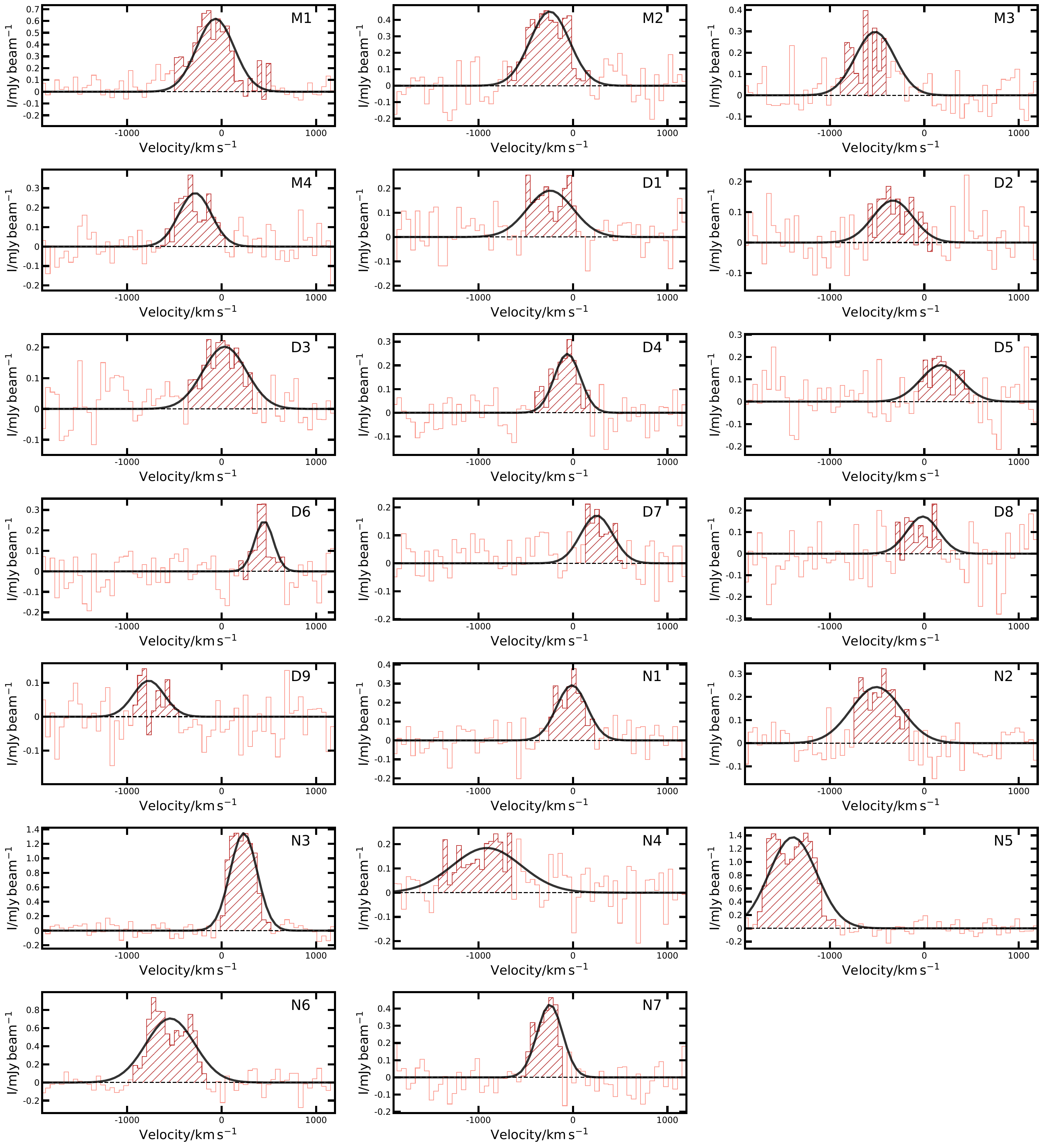}
\caption{\textbf{The CO(3-2) spectral line of detected cluster galaxies.} Each panel shows the CO(3-2) line emission within 1 $\times$ FWHM of the synthesized beam centred at the galaxy. The shaded region is the velocity range. The black solid line is the Gaussian fit to the emission line.}\label{efigure6}
\end{extfigure}

\begin{extfigure}
\centering
\includegraphics[width=1\textwidth]{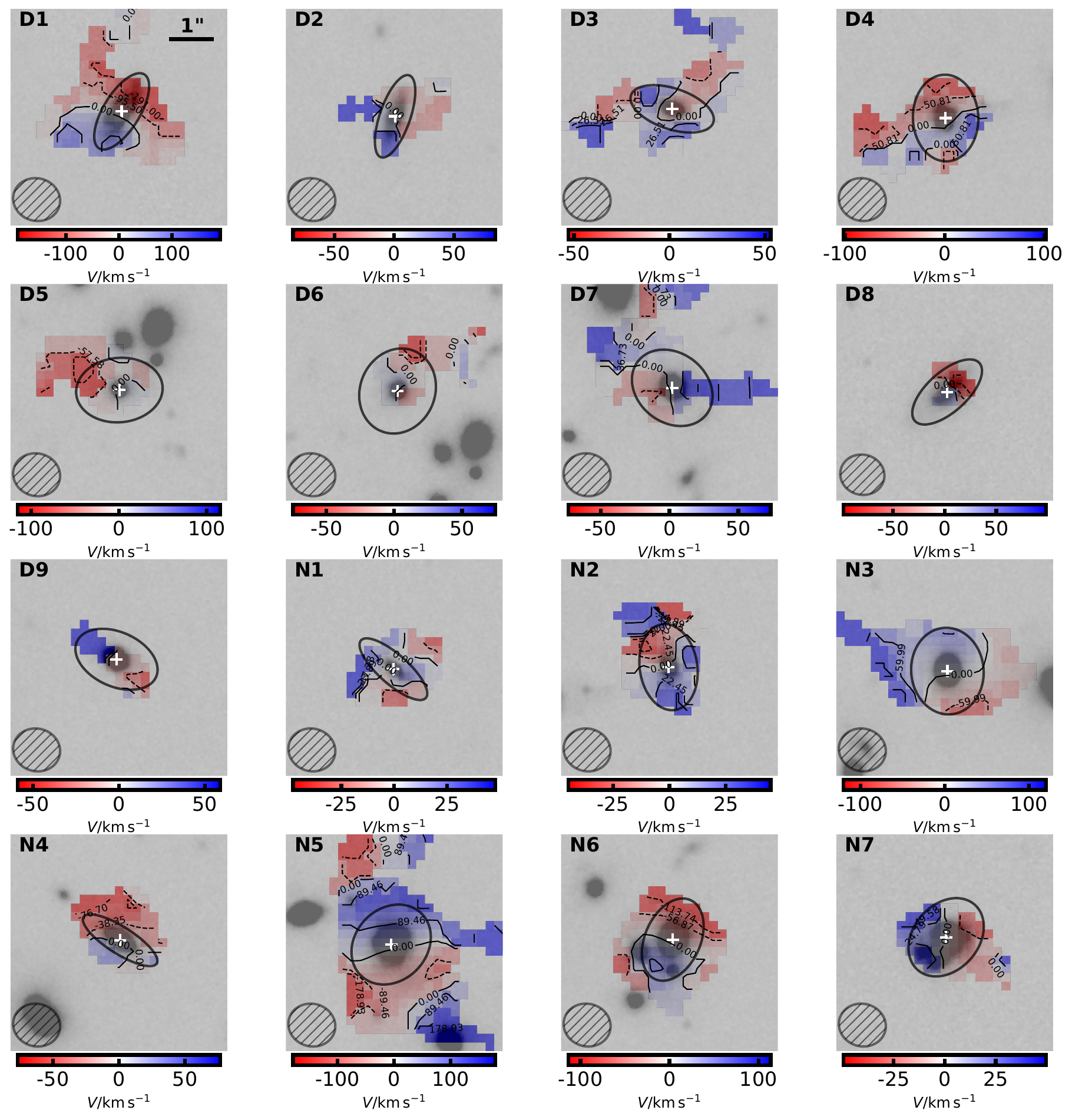}
\caption{\textbf{The velocity maps on the F444W images for the CO(3-2) detected cluster member galaxies.} In each panel, we show the moment-1 maps for the normal and disturbed galaxies. The lower left ellipse denotes the synthesized beam size of ALMA observations and galaxy IDs are labeled in the upper left. The black ellipses at centre denote the shapes of the stellar component from \textsc{PetroFit}. The white plus shows the galaxy centre. The contours and the inline numbers show the velocity gradient. All panels are in size of $5''\times5''$. The scale bar is shown in the upper right of the first panel. }\label{efigure7}
\end{extfigure}

\begin{extfigure}
\centering
\includegraphics[width=0.95\textwidth]{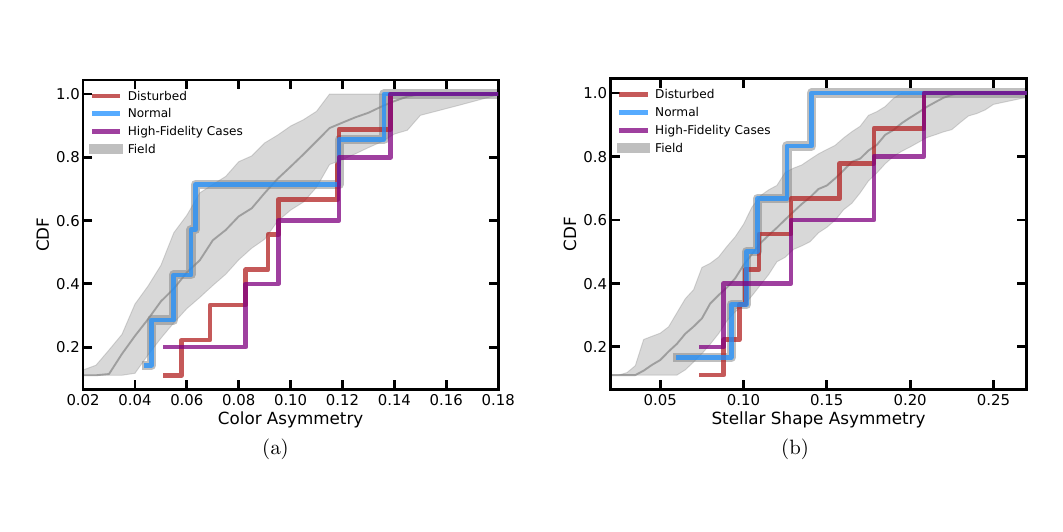}
\caption{\textbf{The cumulative distribution function of the F150W-F277W colour asymmetry and stellar shape asymmetry for galaxies of different categories.} (a) shows the CDF of the color asymmetry for the disturbed galaxies (red), normal galaxies (blue), the high-fidelity RPS galaxies (purple) and field control sample (at the same redshift with matched stellar mass distribution of the disturbed galaxies; grey). The grey shaded region shows the $1\sigma$ bootstrapped errors of the control sample. (b) shows the CDF of stellar shape asymmetry for different groups of galaxies with denotation the same as panel (a).  }\label{efigure8}
\end{extfigure}

\begin{extfigure}
\includegraphics[width=1\textwidth]{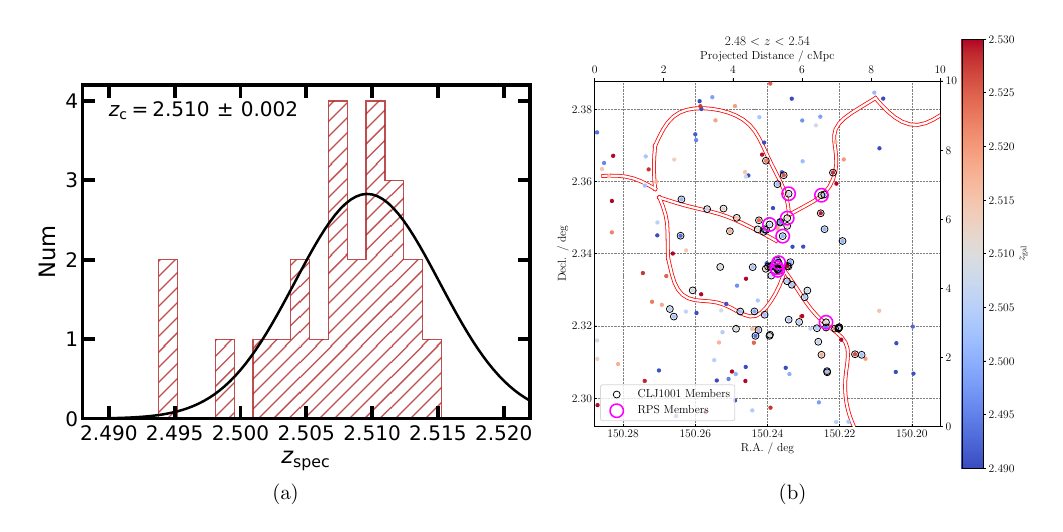} 
\centering

\caption{\textbf{The redshift distribution of spectroscopically confirmed cluster members in CLJ1001 and the reconstructed large-scale structures around CLJ1001.} 
(a): the red histogram shows the redshift distribution of the spectroscopically confirmed cluster members and the solid line denotes the best-fitted Gaussian distribution. The text shows the best-fitted redshift of CLJ1001 with $1\sigma$ standard deviation. (b): the red lines are the reconstructed filaments around CLJ1001. The black-edged points are the confirmed cluster members (from refs.~\cite{Wang+16,Wang+18,Sun:2024} and this work), the magenta circles highlight the disturbed galaxies and background points are the photometric sources from ref.~\cite{Shuntov+2025}.}\label{efigure9}
\end{extfigure}


\begin{table} 
\moveleft 0.9cm \vbox{
\centering
\caption{\textbf{Physical Properties of the CO(3-2)-detected Cluster Members in CLJ1001.}
\footnotetext[a]{the redshift of the CO(3-2) spectral lines from the Gaussian fitting (Extended Data Figure 6). The typical error is $\lesssim  0.001$.}
\footnotetext[b]{the aperture photometry on CO(3-2) moment-0 map within convolved stellar effective radius.} 
\footnotetext[c]{the global star-formation rate from SED fitting.}
\footnotetext[d]{the unconvolved effective radius from S${\rm \acute{e}}$rsic model in JWST/F444W image with the convolved effective radius in the brackets.}} \label{etab1}
\begin{tabular}{lllllllll} 
\\
\hline
Galaxy & Ra  & Dec & $z_{\rm CO(3-2)}$$^a$ & $\log M_\star$&$\log L_{\rm CO(3-2),eff}$$^b$ &$\log\rm SFR$$^c$ &${r}_{\rm gas}$ & ${r_{\rm eff,\star}}$$^d$\\
&   &  &  & ($M_\odot$)&(${\rm K\,km\,s^{-1}pc^{2}}$) &($M_\odot{\rm yr^{-1}}$) &($''$) & ($''$)\\
\hline
M1    & 150.22026   & 2.31932  & 2.509 &... & ...&...&... & ...\\
M2    & 150.23913   & 2.33633  & 2.507 &... & ...&...&... & ...\\
M3    & 150.23421   & 2.33648  & 2.504 &... & ...&...&... & ...\\
M4    & 150.23647   & 2.34871  & 2.507 &... & ...&...&... & ...\\
D1    & 150.23412   & 2.35662  & 2.507 &10.40$\pm$0.17&9.64$\pm$0.08 & 2.19$\pm$0.15&0.89$\pm$0.08 & 0.44
(0.67)\\
D2    & 150.23942   & 2.34809  & 2.506 &10.55$\pm$0.06&9.32$\pm$0.17 & 1.70$\pm$0.10&0.86$\pm$0.14 & 0.42
(0.65)\\
D3    & 150.23452   & 2.34986  & 2.510 &10.47$\pm$0.06&9.50$\pm$0.10 & 1.68$\pm$0.13&0.98$\pm$0.12 & 0.18
(0.58)\\
D4    & 150.22502   & 2.35619  & 2.509 &9.98$\pm$0.04&9.48$\pm$0.10 & 2.04$\pm$0.02&1.07$\pm$0.09 & 0.23
(0.67)\\
D5    & 150.23715   & 2.33538  & 2.512 &10.54$\pm$0.05&9.21$\pm$0.20 & 1.62$\pm$0.08&1.20$\pm$0.22 & 0.12
(0.59)\\
D6    & 150.23741   & 2.33609  & 2.515 &10.56$\pm$0.05&9.17$\pm$0.22& 1.75$\pm$0.11&1.15$\pm$0.23 & 0.13
(0.61)\\
D7    & 150.23691   & 2.33750  & 2.513 &10.89$\pm$0.05&9.19$\pm$0.16 & 2.99$\pm$0.07&1.40$\pm$0.28 & 0.35
(0.65)\\
D8    & 150.22377   & 2.32100  & 2.510 &10.64$\pm$0.10&9.45$\pm$0.21 & 1.75$\pm$0.09&0.49$\pm$0.09 & 0.62
(0.69)\\
D9    & 150.23578   & 2.34486  & 2.501 &10.62$\pm$0.08&8.86$\pm$0.31 & 1.86$\pm$0.13&0.91$\pm$0.20 & 0.12
(0.58)\\
N1    & 150.24272   & 2.34674  & 2.510 &10.30$\pm$0.06&9.55$\pm$0.10 & 1.47$\pm$0.11&0.64$\pm$0.05 & 0.21
(0.59)\\
N2    & 150.22426   & 2.35642  & 2.504 &10.53$\pm$0.12&9.59$\pm$0.08& 1.73$\pm$0.14&0.60$\pm$0.04 & 0.22
(0.59)\\
N3    & 150.23986   & 2.33645  & 2.513 &11.04$\pm$0.07&10.12$\pm$0.02& 2.21$\pm$0.12&0.61$\pm$0.01 & 0.17
(0.59)\\
N4    & 150.23864   & 2.33682  & 2.499 &11.03$\pm$0.04&9.51$\pm$0.11& 2.06$\pm$0.15&0.62$\pm$0.06 & 0.20
(0.62)\\
N5    & 150.23728   & 2.33813  & 2.494 &11.30$\pm$0.05&10.31$\pm$0.01& 2.33$\pm$0.10&0.62$\pm$0.01 & 0.33
(0.59)\\
N6    & 150.23691   & 2.33577  & 2.504 &11.02$\pm$0.08&10.03$\pm$0.03& 2.37$\pm$0.14&0.62$\pm$0.01 & 0.21
(0.59)\\
N7    & 150.22891   & 2.32984  & 2.507 &10.99$\pm$0.06&10.15$\pm$0.08 & 2.37$\pm$0.11&0.69$\pm$0.07 & 0.54
(0.68)\\
\hline
\end{tabular}
}
\end{table}

\end{document}